\documentclass[letterpaper,12pt]{article}   
\usepackage{osajnl2} 
\usepackage{amsmath,amssymb}             

\begin{document}

\title{Spatial amplitude and phase modulation using commercial twisted nematic LCDs}
\author{E.~G. van Putten$^{*}$, I.~M. Vellekoop and A.~P. Mosk}
\address{Complex Photonic Systems, Faculty of Science and Technology\\
and MESA$^+$ Institute for Nanotechnology, University of Twente\\
P.O.Box 217, 7500 AE Enschede, The Netherlands}
\address{$^*$Corresponding author: E.G.vanPutten@utwente.nl}

\begin{abstract}
    We present a method for full spatial phase and amplitude control of a laser beam using a twisted nematic liquid crystal display combined with a spatial filter. By spatial filtering we combine four neighboring pixels into one superpixel. At each superpixel we are able to independently modulate the phase and the amplitude of light. We demonstrate experimentally the independent phase and amplitude modulation using this novel technique. Our technique does not impose special requirements on the spatial light modulator and allows precise control of fields even with imperfect modulators.
\end{abstract}


\maketitle 


        \section{Introduction}
        Several exciting new applications of optics rely on spatial phase and amplitude control of light. In adaptive optics, light is modulated to correct for abberations in a variety of optical systems such as the human eye \cite{LeGargasson2001} or the atmosphere \cite{Beckers1993}. Digital holography is another area of optics which relies on the spatial modulation of light. Holographic data storage \cite{Psaltis1998}, 3D display technology \cite{Travis1997}, and diffractive optical elements \cite{Davis2001} are just a few examples of the many exciting topics in digital holography that require a very precise modulation of light. Very recently, it was shown that light can be focussed through and inside opaque strongly scattering materials by sending in spatially shaped wavefronts \cite{Vellekoop2007,Vellekoop2008aa}.

        Liquid crystal spatial light modulators (LC SLMs) are used in many cases where modulation of light is required because of their high optical efficiency, high number of degrees of freedom, and wide availability. A LC SLM can modulate the phase and the amplitude of light at typical refresh rates of $\sim$60~Hz and at resolutions in the order of $10^6$ pixels. Aside from their advantages, commercially available SLMs have some limitations. In most LC SLMs the modulation of phase is coupled to a modulation of the polarization. This makes it hard to independently modulate the phase or the amplitude.

        Several techniques have been proposed to achieve independent spatial phase and amplitude control using a SLM. Examples are setups where two SLMs are used to balance out the cross modulation \cite{Kelly1998aa,Neto1996aa}, double pass configurations where the light propagates twice through the same SLM \cite{Dou1996aa,Chavali2007aa}, and double pixel setups which divide the encoding of complex values over two neighboring pixels \cite{Bagnoud2004,Arrizon2003,Birch2000}. Each of these techniques has its specific limitations. The use of two SLMs in one setup \cite{Kelly1998aa,Neto1996aa} requires a sensitive alignment and valuable space. Double pass configurations \cite{Dou1996aa,Chavali2007aa} double the phase modulation capacities of the SLM, usually delivering a full $2\pi$ phase shift. Unfortunately though, to keep the amplitude constant, the SLM has to work in a mostly-phase condition which requires light of a specific elliptical polarization. Double pixel setups \cite{Bagnoud2004,Arrizon2003,Birch2000} also require special modulation properties like amplitude-only modulation \cite{Birch2000}, phase-only modulation \cite{Bagnoud2004}, or a phase modulation range of $2\pi$ \cite{Arrizon2003}. These properties are usually achieved by sending in light under certain specific polarizations. Small deviations from these polarizations already modify the modulation properties, leading to drift in the achieved amplitudes and phases.

        We propose a novel modulation technique to achieve amplitude and phase modulation of light using a SLM together with a spatial filter. This technique does not pose special requirements on the SLM or the incoming light, making our technique more widely applicable than previous techniques. In our experiments we used a HoloEye LC-R 2500 twisted nematic liquid crystal display (TN-LCD) which has $1024$x$768$ pixels ($19~\mu$m square in size) and which can modulate with 8-bit accuracy (256 voltage levels). This SLM has a strong cross-modulation between the phase and the polarization of the light. We demonstrate the use of this SLM to independently modulate phase and amplitude.

        \section{Cross-modulation in TN-LCDs}
        \begin{figure}[t]
            \centering
            \includegraphics[width=7 cm]{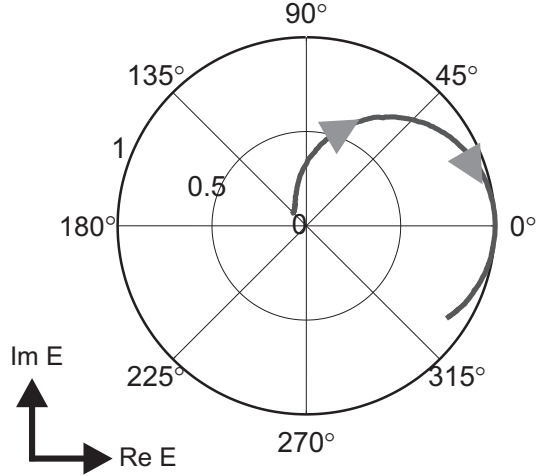}
            \caption{(Color online) Experimental modulation plot for the SLM, illuminated with vertical polarized light. After modulation, the light with a horizontal polarization is detected. This plot shows the amplitude versus the phase modulation. The arbitrary phase reference is chosen such that the maximum amplitude coincides with the a zero phase modulation. The modulation voltage increases in the direction of the gray arrows.}\label{fig:SinglePixelResponse}
        \end{figure}
        Twisted nematic liquid crystal displays (TN-LCDs) owe their name to the helical structure formed by the molecules inside the LCD cells, in absence of an electric field. When a voltage is applied over the edges of the LCD cell, the molecules are aligned to the electric field and the twisted structure disappears. By changing the voltage over the cells, the internal structure is changed continuously from a twisted to a straight alignment of the molecules, affecting both the phase and the polarization of the reflected light.

        To determine the relation between the applied voltage and the modulation of the light, the modulation curve of our SLM, we used a diffractive technique where we placed a binary grating with a 50\% duty cycle on the display, comparable with the technique described in \cite{Zhang1994aa}. Using a polarizing beam splitter we illuminated the SLM with vertically polarized light and detected the horizontally polarized component of the modulated light. We measured the intensity in the 0$^{th}$ order of the far-field diffraction pattern for different contrast levels of the grating. The contrast difference is achieved by keeping the notches of the grating steady at a constant value while we cycled the voltage values of the rules of the grating over the complete voltage range in 256 steps. We determined the real and imaginary parts of the field as a function of the voltage level encoded on the pixels, by comparing the measured $0^{th}$ order intensity for different pixel values on both the notches and the rules of the grating. The results of these measurements are plotted in Fig.~\ref{fig:SinglePixelResponse}. When no voltage is applied there is a $0.75\pi$ phaseshift and the amplitude is close to zero. With increasing voltage, the phase is modulated from $0.75\pi$ back to $0$ where the amplitude is at a maximum. If the voltage is increased further to the maximal voltage, the phase is modulated up to $-0.62\pi$ while the amplitude decreases slightly. The deviation of the modulation curve from a circle centered around the origin, is a clear indication of cross-modulation.

        \section{Decoupling of phase and amplitude by combining multiple pixels}
        \label{sec:DecouplingPhaseAmplitude}        
        \begin{figure}[t]
            \centering
            \includegraphics[width=8.4 cm]{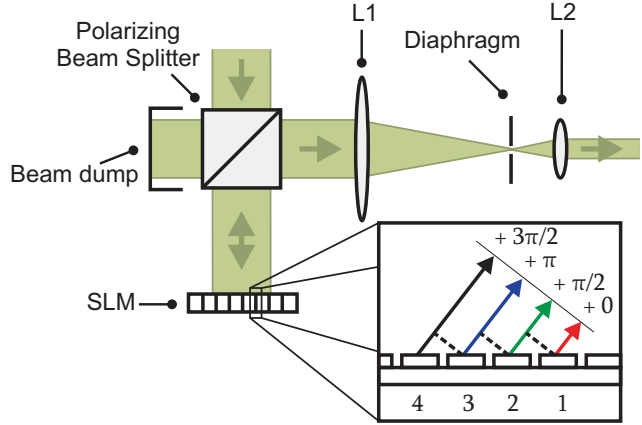}
            \caption{(Color online) Experimental setup to decouple phase and amplitude modulation. Four neighboring pixels, pixels 1, 2, 3, and 4, are combined to one superpixel which can modulate one complex value. The modulated light is focused by a lens, L1, onto a diaphragm placed in the focal plane. The focal length of L1 is $200$~mm. The diaphragm is positioned such that it only transmits light under a certain angle. This angle is chosen so that, in the horizontal dimension, two neighboring pixels are exactly $\pi/2$ out of phase with each other, as can be seen from the inset. Behind the diaphragm, the light is collimated by a lens, L2.}\label{fig:Setup}
        \end{figure}
        
        \begin{figure}[t]
            \centering
            \includegraphics[width=7 cm]{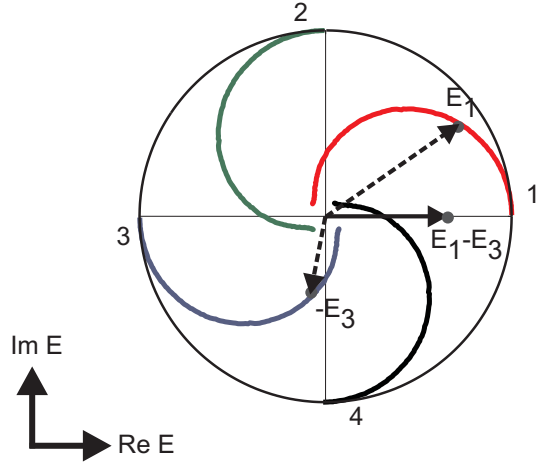}
            \caption{(Color online) The electric field of a superpixel is superposed out of four distinct neighboring pixels, which phases are all shifted over $\pi/2$ with respect to each other. In this figure we show a limited modulation range of the pixels 1, 2, 3, and 4. The numbers outside the plot relate the curves to the pixel numbers. By choosing the right pixel values for the different pixels, every complex value can be synthesized by the superpixel. In the figure pixel 1 and 3 are used to modulate a value on the real axis. The electric fields $E_1$ and $E_3$ are chosen such that their imaginary parts cancel and $E_1-E_3$ lies on the real axis. Pixels 1 and 3 are used in a similar way to modulate any value on the imaginary axis.}\label{fig:ResponseFourPixels}
        \end{figure}
        By combining four neighboring pixels into one superpixel, we decouple the modulation of the phase and the amplitude, thereby achieving full independent phase and amplitude control of light. Any arbitrary complex amplitude can be synthesized even though each individual pixel on the SLM has a limited complex response (see Fig.~\ref{fig:SinglePixelResponse}). This technique is based on the work done by Birch et al. \cite{Birch2000,Birch2001aa} where they combined two neighboring pixels of a amplitude-only SLM, to achieve full independent phase and amplitude modulation. By using four pixels instead of two, our technique no longer depends on a special modulation curve. There are only two requirements on the modulation curve. First of all, the real part of the modulation curve should be an increasing function of the applied voltage. Secondly, for each imaginary value on the modulation curve where the derivative is non-zero, there should exist at least two different solutions. For every known SLM the modulation curve, or a suitable subset, follows these two requirements.

        The setup used to decouple the phase and amplitude modulation is shown in Fig.~\ref{fig:Setup}. A monochromatic beam of light at a wavelength of $532$~nm is incident normal to the SLM surface. The modulated light is reflected from the SLM. We choose an observation plane at an angle at which the contribution of each neighboring pixel is $\pi/2$ out of phase, as is seen in the inset. The fields from the individual pixels are superposed using a spatial filter. By choosing the correct combination of pixel voltages, any complex value within a square (-1..1),(-i..i) can be synthesized.

        To construct a complex value $f = g + ih$, we use the first and third pixel of a superpixel to construct $g$ and the second and the fourth pixel to construct $h$. The voltage for the first and third pixel, $V_1$ and $V_3$, are chosen such that the imaginary parts of the modulated fields are equal and the difference between the real parts of the modulated fields is equal to $g$ (a detailed discussion about how to find the voltages can be found in Appendix~\ref{App:howToFindVoltages}). The fields modulated by the two pixels are then
        \begin{eqnarray}
            E_{1} &=& E_{1r} + i\Delta,\\ \label{eq:E_1}
            E_{3} &=& E_{3r} + i\Delta,\\ \label{eq:E_3}
            E_{1r} - E_{3r} &=& g, \label{eq:E_1-E_3}
        \end{eqnarray}
        where $E_{1r}$ and $E_{3r}$ are the real parts of the fields modulated by pixel $1$ and $3$, and $\Delta$ is the imaginary term of the modulated fields. In the plane of reconstruction, the two pixels are $\pi$ out of phase. Therefore the imaginary parts of the modulated fields cancel each other while the real parts are subtracted and produce a non-zero real value equal to $g$. A geometrical representation of this construction is given in Fig.~\ref{fig:ResponseFourPixels}. We construct $h$ in a similar way using the second and the fourth pixel. The fields modulated by the these two pixels are then chosen to be
        \begin{eqnarray}
            E_{2} &=& E_{2r} + i\epsilon,\\ \label{eq:E_2}
            E_{4} &=& E_{4r} + i\epsilon,\\ \label{eq:E_4}
            E_{2r} - E_{3r} &=& h. \label{eq:E_2-E_4}
        \end{eqnarray}
        The $\pi/2$ phase shift between all the neighboring pixels in a superpixel causes that the reconstructed values $g$ and $h$ are $\pi/2$ out of phase. The total reconstructed field is therefore $g + ih$ which is exactly the complex value $f$ we wanted to modulate.

        Instead of a single value $f$, we use the SLM to construct a two dimensional function $f(x,y) = g(x,y) + ih(x,y)$. We will calculate the reconstructed field in analogy to the analysis given in Birch et al. \cite{Birch2001aa}. When we modulate the function $f(x,y)$ the field at the SLM surface is described by
        \begin{eqnarray}
        \nonumber
        	f_s (x,y)&=&
    		\sum_{m,n=-\infty}^{\infty}{
                    \delta(y - m a/4)} \Big[ \delta(x-na) E_1(x,y)
        \\ \nonumber
                 && ~~~~~~~~~~~~~~~~~~~~~~~~
                 + \delta(x-1/4a-na) E_2(x-1/4a,y)
        \\ \nonumber
                 && ~~~~~~~~~~~~~~~~~~~~~~~~
                 + \delta(x-1/2a-na) E_3(x-1/2a,y)
        \\
                 && ~~~~~~~~~~~~~~~~~~~~~~~~
                 + \delta(x-3/4a-na) E_4(x-1/4a,y) \Big],
        \label{eq:FieldOnSLM}
        \end{eqnarray}
        where the shape of the pixels is simplified to a delta function response. The functions $E_1$, $E_2$, $E_3$, and $E_4$ are constructed such that for each superpixel coordinate (na,ma/4) they agree with Eq.~\ref{eq:E_1}--\ref{eq:E_2-E_4}. A lens $L_1$ positioned behind our SLM Fourier transforms the field. To find the field in the focal plane, we calculate the Fourier transform of Eq.~\ref{eq:FieldOnSLM} to the spatial frequencies $\omega_x$ and $\omega_y$ and recenter the coordinate system to the center of the first diffraction mode by defining $\Omega_x \equiv \omega_x - \frac{2\pi}{a}$:
        \begin{eqnarray}
        \nonumber
            F_s(\Omega_x,\omega_y) &=&
                \frac{1}{2\pi}
                \sum_{k,l=-\infty}^{\infty}{}
                    \Bigg\{~~~~
                    \tilde{E}_{1r}\left( \Omega_{xk}, \omega_{yl}\right)
                    - \tilde{E}_{3r}\left( \Omega_{xk}, \omega_{yl}\right)
                    e^{-i\frac{a\Omega_{x}}{2}}
        \\ \nonumber
        && ~~~~~~~~~~~~~+i\left[
                    \tilde{E}_{2r}\left( \Omega_{xk}, \omega_{yl}\right)
                    - \tilde{E}_{4r}\left( \Omega_{xk}, \omega_{yl} \right)
                    e^{-i \frac{a\Omega_x}{2}}
                 \right] e^{-i\frac{a\Omega_x}{4}}
        \\
        && ~~~~~~~~~~~~~+i\left[
                \tilde{\Delta}\left( \Omega_{xk}, \omega_{yl}\right) +
                \tilde{\epsilon}\left( \Omega_{xk}, \omega_{yl}\right) e^{-i\frac{a\Omega_x}{4}}
                \right]
                \left[1-e^{-i\frac{a\Omega_x}{2}} \right] \Bigg\}
        \label{eq:FilteredFourier}
        \end{eqnarray}
        with $\Omega_{xk} \equiv \Omega_x - 2\pi k/a$ and $\omega_{yl} \equiv \omega_y - 8\pi l/a$. From this equation we see that at $\Omega_x = 0$, the $\tilde{\Delta}$ and $\tilde{\epsilon}$ terms completely vanish and only the terms $\tilde{E}_{1r}-\tilde{E}_{3r}+i(\tilde{E}_{2r}-\tilde{E}_{4r}) = \tilde{g} + i\tilde{h}$ survive, which is exactly the Fourier transform of the original function written to the SLM. At non-zero spatial frequencies the $\tilde{\Delta}$ and $\tilde{\epsilon}$ terms appear and introduce an error which increases with spatial frequency. To reduce these errors and cut out unwanted diffraction orders, we place an diaphragm of width $2\pi (\lambda f w)/a$ around $(\Omega_x,\omega_y) = (0,0)$, with $\lambda$ the wavelength of the light, $f$ the focal length of lens $L_1$ and $w$ the relative width of the diaphragm. The relative width can be chosen $w=1$ for maximal resolution, but can also be chosen smaller. We note that in the original analysis by Birch et al. \cite{Birch2001aa} all terms inside the summation, except one, were neglected. However, at the SLM pixels, light may diffract into different orders. Therefore we do not make this approximation and include the summation over all orders into our calculation.

        \begin{figure}
            \centering
            \includegraphics[width=8 cm]{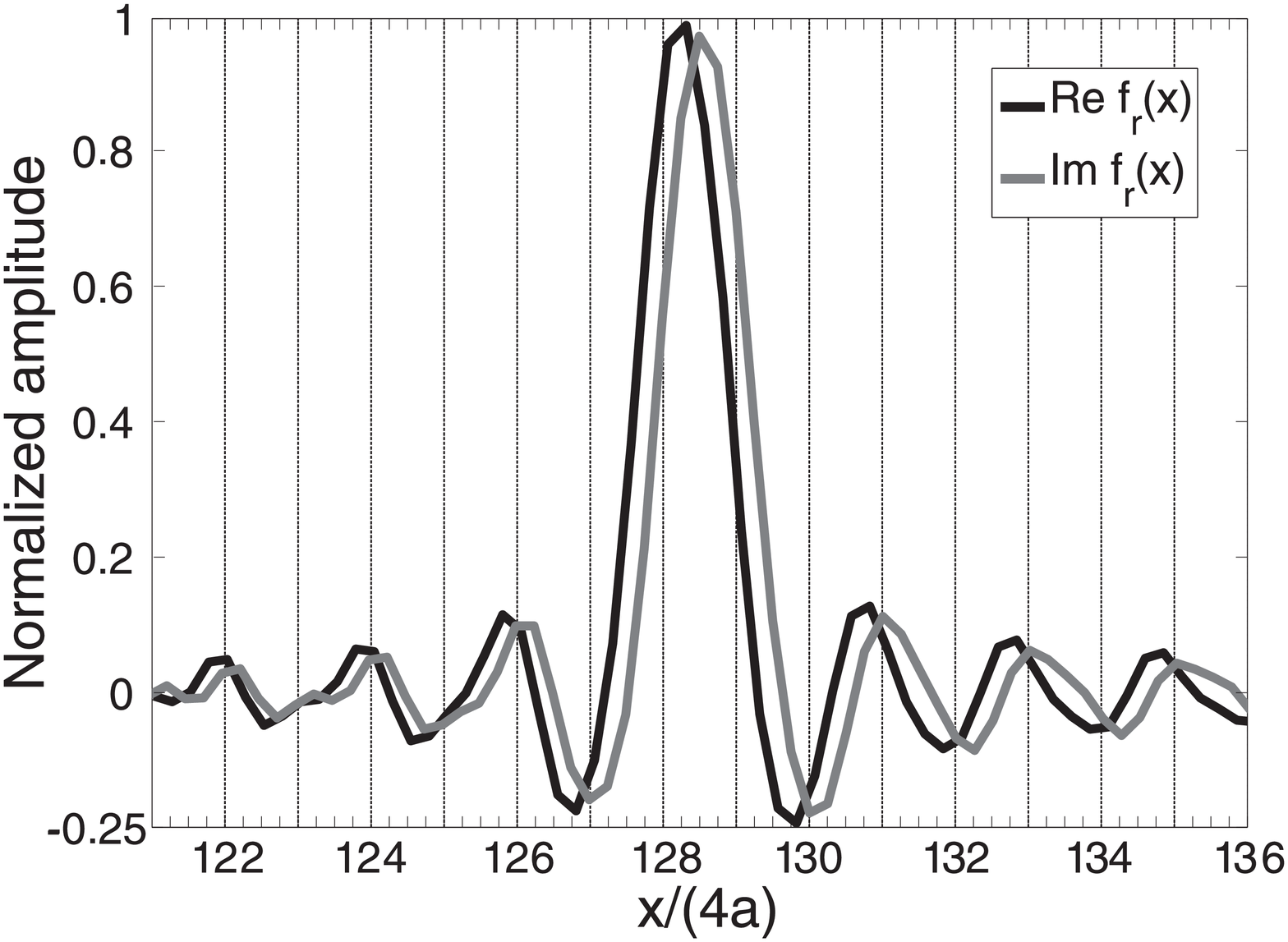}[t]
            \caption{(Color online) Calculation of the reconstructed field in the x-direction for the case where the $128^{th}$ superpixel is set to $1+i$ while the other superpixels are set to 0. The vertical lines represent the borders of the superpixels. The reconstructed real and imaginary part are spatially separated by $a/4$.}\label{fig:ReconstructedField}
        \end{figure}
        A second lens $L_2$ transforms the filtered field at the diaphragm into the reconstructed field. We find this reconstructed field by taking the inverse Fourier transform of Eq. \ref{eq:FilteredFourier}:
        \begin{eqnarray}
        \nonumber
        	f_r (x,y)&=&
    		\sum_{m,n=-\infty}^{\infty}{
                    \delta(y - m a/4)} \Big[ \delta(x-na) E_1(x,y)
        \\ \nonumber
                 && ~~~~~~~~~~~~~~~~~~~~~~~~
                 - \delta(x-1/2a-na) E_3(x-1/2a,y)
        \\ \nonumber
                 && ~~~~~~~~~~~~~~~~~~~~~~~~
                 + i\delta(x-1/4a-na) E_2(x-1/4a,y)
        \\ \nonumber
                 && ~~~~~~~~~~~~~~~~~~~~~~~~
                 - i\delta(x-3/4a-na) E_4(x-1/4a,y) \Big]
        \\
                 && ~~~~~~~~~~~~~~~~~~~~~~~~
                 \otimes \frac{2w}{a^2} \text{sinc}\left(\frac{xw}{a},\frac{4y}{a}\right).
        \label{eq:ReconstructedField}
        \end{eqnarray}
        In the reconstructed field, the pixels of a superpixel are spatially separated in the $x$-direction, a limitation that is present in every multipixel technique. Due to the spatial filter, the fields modulated by different subpixels are spatially broadened which effectively averages them. In Fig.~\ref{fig:ReconstructedField} the reconstructed field in the x-direction is shown for the case where the $128^{th}$ superpixel is set to $1+i$ while the other superpixels are set to 0. The real and imaginary part of the reconstructed superpixel are spatially separated by $a/4$. Decreasing the size of the spatial filter, $w$, results in an improved quality of the reconstructed field by cutting out the $\tilde{\Delta}$ and $\tilde{\epsilon}$ terms in Eq.~\ref{eq:FilteredFourier}. This quality improvement happens at the cost of spatial resolution, as for small $w$ the sinc-function in Eq.~\ref{eq:ReconstructedField} is broadened.

        \section{Experimental demonstration of modulation technique}
        \begin{figure}[t]
            \centering
            \includegraphics[width=7 cm]{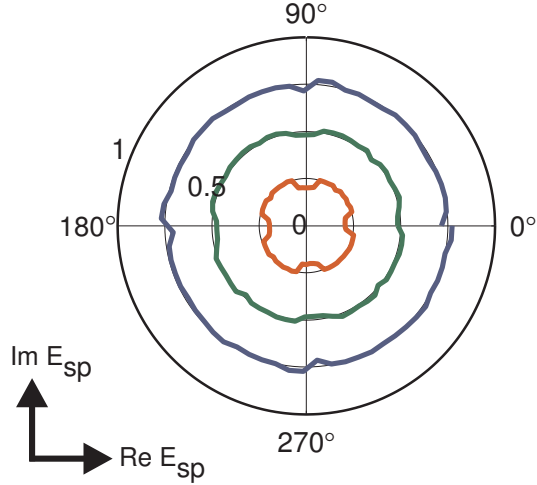}
            \caption{(Color online) Results of the measurements to demonstrate the independent phase and amplitude modulation. Using the superpixels, we synthesized a plane wave of which we varied the amplitude and the phase. In this plot see the relative amplitude $A/\sqrt{I_0}$ from intensity measurements as a function of the programmed phase. The intensity $I_0 = 19.7\cdot10^{3}$ counts/seconds. The relative amplitudes are set to 0.25, 0.5, and 0.75, corresponding to the red, green, and blue line respectively.}\label{fig:PolarPlotPhaseAmplitude}
        \end{figure}

        \begin{figure}[t]
            \centering
            \includegraphics[width=8 cm]{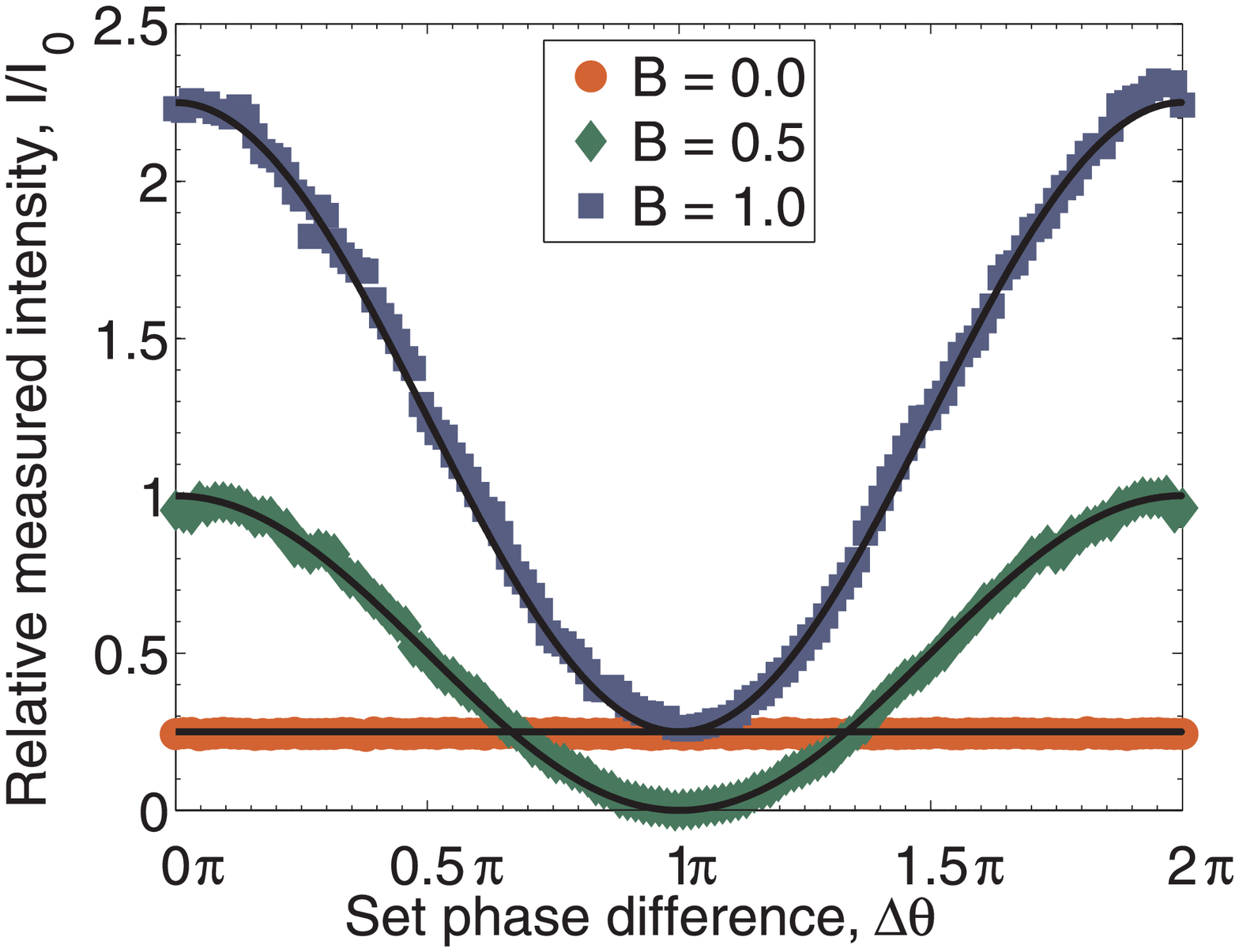}
            \caption{(Color online) Results of the measurements to demonstrate that the set phase difference corresponds with the actual phase difference. In this plot we show the relative measured intensity of a grating interference experiment. The intensities are plotted as a function of the set phase difference $\Delta\theta$ and are reference to $I_0 = 4.56\cdot10^{3}$ counts/seconds. The experiment was done for multiple amplitude differences between the rules and the notches of the grating. The solid lines represent the expected intensities.}\label{fig:Plot_IntensityVsPhase}
        \end{figure}
        We tested the modulation technique with the setup depicted in Fig. \ref{fig:Setup}. We placed an extra lens and a CCD camera behind the spatial filter to detect the modulated light in the far-field.

        First we demonstrate the decoupling of the phase and amplitude modulation. We encoded the same amplitudes and phases on all superpixels of the SLM, forming a plane light wave. The plane wave was then focused onto the CCD camera. We measured the intensity in the focus while changing the phase and amplitude of the plane wave. In Fig. \ref{fig:PolarPlotPhaseAmplitude} the relative amplitudes from the intensity measurements are plotted against the phase of the plane wave. For different amplitude modulations, the phase is cycled from 0 to $2\pi$. It is seen from the results, that we modulate the phase while keeping the amplitude of the light constant within $2.5~\%$.

        Next we show that the set phase modulation is in agreement with the measured phase modulation. We placed a binary grating with a 50\% duty cycle on the SLM. We modified the contrast by keeping one part of the grating constant at a relative amplitude, $A$, of 0.5 while changing the relative amplitude of the other part, $B$ between 0 and 1. For each value of $B$ we changed the phase difference between the two parts of the grating, $\Delta\theta$, from 0 to 2$\pi$. Because of interference, the recorded intensity, I, at the CCD camera changes as a function of A,B, and $\Delta\theta$ in the following way:
        \begin{equation}
            I = A^2 + B^2 + 2AB\cos(\Delta\theta).
        \label{eq:IntensityWithInterference}
        \end{equation}
        In Fig. \ref{fig:Plot_IntensityVsPhase} the results are presented for three different values of B. Both the amplitude and the phase of the measured oscillations match the theory well, demonstrating an independent phase modulation in agreement with the set phase difference. With the new modulation technique we are able to modulate the phase of the light from $0$ to $2\pi$, even though each individual pixel is not able to modulate a full $2\pi$ phase shift (see Fig. \ref{fig:SinglePixelResponse}).

        \begin{figure}[t]
            \centering
            \includegraphics[width=8.4 cm]{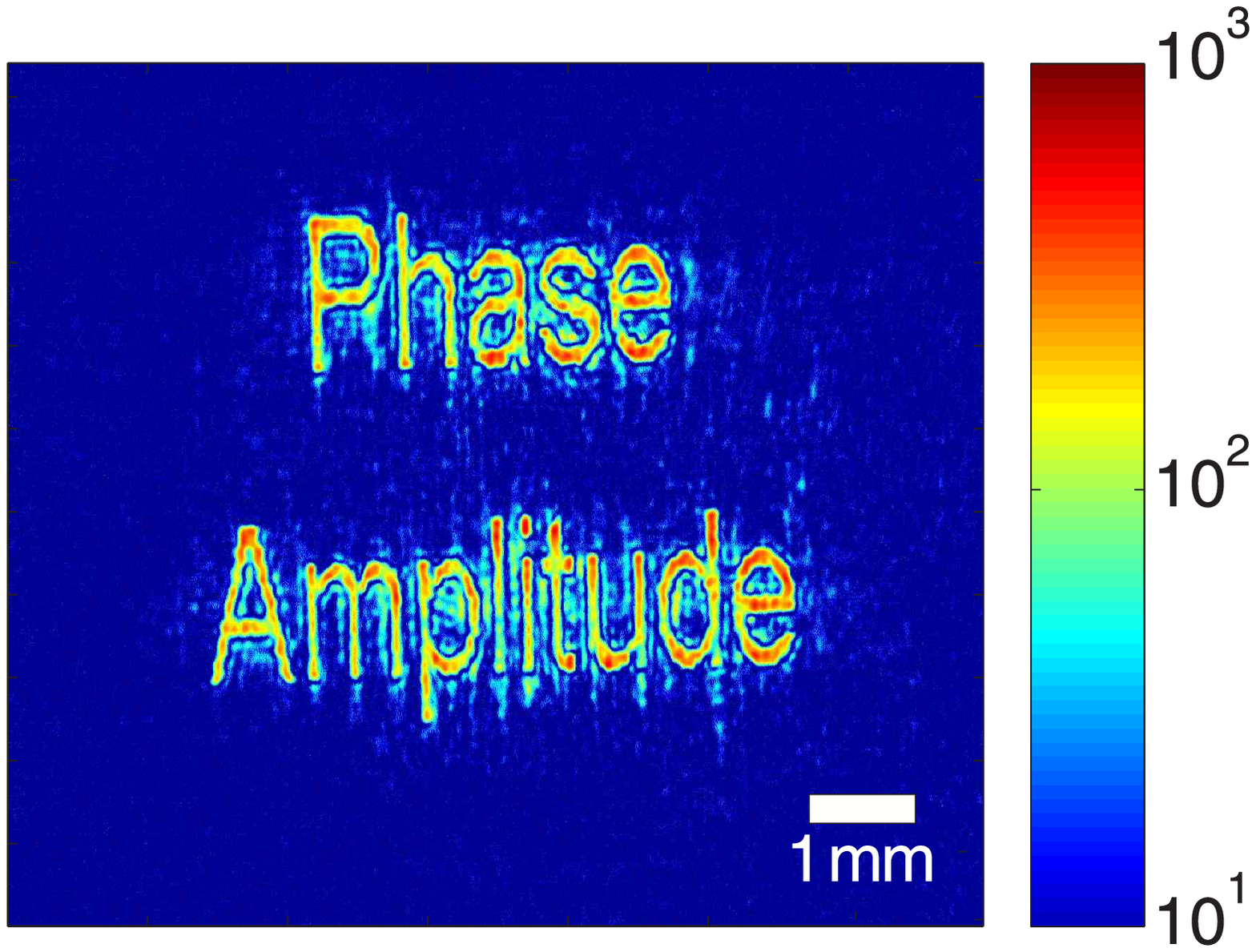}
            \caption{(Color online) Demonstration of complex modulation using the novel modulation technique. In this experiment we encoded the Fourier transform of an image containing the two words \textit{Phase} and \textit{Amplitude} onto the SLM and reconstructed the image in the far field. This plot shows the measured intensity in the far field on a logarithmic scale. The intensities are in counts/seconds.}\label{fig:FFTPhaseAndAmplitude}
        \end{figure}

        Finally we demonstrate complex spatial modulation of light. We encoded the Fourier transform of an image onto the SLM. The original image is reconstructed in the far-field and was recorded by the CCD camera. In Fig. \ref{fig:FFTPhaseAndAmplitude} the reconstructed intensity is shown on a logarithmic scale. The letters in the image stand out with high contrast. Around the letters, small areas containing noise are seen. The relative intensity of the noise is smaller than 10$\%$.

        \section{Conclusions}
        We have presented a novel approach to achieve full independent spatial phase and amplitude control using a twisted nematic liquid crystal display together with a spatial filter. Our novel modulation scheme combines four neighboring pixels into one superpixel. Each superpixel is able to independently generate any complex amplitude value. We demonstrated that we were able to freely modulate the phase over a range of 2$\pi$ while keeping the amplitude of the light constant within 2.5$\%$. Furthermore we showed that we could achieve a spatially complex modulation of the light with a high resolution.

        The presented modulation technique can be used with any SLM. No special modulation response is needed, therefore the illumination light can have linear polarization. This polarization makes a relative simple calibration possible. Using linearly polarized light, the setup can easily be expanded to control two independent polarizations with two SLMs.

        \section*{Acknowledgements}
        The authors thank Ad Lagendijk and Willem Vos for support and for valuable discussions. This work is part of the research programme of the `Stichting voor Fundamenteel Onderzoek der Materie (FOM)', which is financially supported by the `Nederlandse Organisatie voor Wetenschappelijk Onderzoek' (NWO). A.P. Mosk is supported by a VIDI grant from NWO.

        \appendix
        \section{How to find the proper pixel voltages}\label{App:howToFindVoltages}
        In this appendix, we prove that for every modulation curve that fulfills the requirements stated in Section~\ref{sec:DecouplingPhaseAmplitude} we can find pixel voltages fulfilling Eq.~\ref{eq:E_1}--\ref{eq:E_2-E_4}. The real and imaginary parts of the field modulated by a pixel, $E_r$ and $E_i$ are functions of the voltage on the pixel. We only use the voltage range for which $E_r(V)$ is a rising function of the applied voltage. By multiplying the modulation curve with a proper phase factor, this range can be maximized. The function $E_r(V)$ can be inverted while $E_i(V)$ has two inverses. We divide the voltage range into two parts, $V<V_c$ and $V>V_c$, where $V_c$ is the voltage for which the derivative of $E_i(V)$ equals $0$. On both ranges we can invert the function $E_i(V)$.

        To modulate a positive real value $g(V_1)$, we use pixels 1 and 3 that are $\pi$ out of phase. On the first pixel we place $V_1>V_c$ and on the third pixel we place $V_3 = V_L(E_i(V_1))$ where $V_L$ is the inverse of $E_i(V)$ for $V<V_c$. The total field modulated by these two pixels is now
        \begin{eqnarray}
        g(V_1) = E_r(V_1) - E_r(V_L(E_i(V_1))).
        \end{eqnarray}
        It can be shown that $g(V_1)$ is an increasing function of $V_1$ and that $g(V_c)=0$. We find the function $V_1(g)$ by inverting $g(V_1)$. To construct a negative real value, the voltage $V_1$ should be chosen $V_1<V_c$. The voltage $V_3$ is then $V_3 = V_H(E_i(V_1))$, where $V_H$ is the inverse of $E_i(V)$ for $V>V_c$. Similar calculations can be done to find the functions $V_2(h)$ and $V_4(h)$ to modulate a positive and negative value $h$ using pixels 2 and 4.

        - As a technical note: we reconfigured the gamma lookup curve of the modulation electronics after the calibration measurements to perform the calculations discussed above. The conversion of amplitude and phase to real and imaginary values is performed in real time in the computer's video hardware.


\end{document}